\documentclass[aps,preprint,prd,showpacs,nofootinbib]{revtex4}

\usepackage{latexsym}
\usepackage{amsmath}
\usepackage{graphicx}
\usepackage{subfigure}
\usepackage{dcolumn}
\usepackage{bm}
\usepackage{amssymb}
\usepackage{color}
\usepackage{float}
\usepackage[colorlinks,linkcolor=magenta,anchorcolor=cyan,citecolor=blue]{hyperref}

\hypersetup{colorlinks=true,
    breaklinks=true,
    pdfstartview=Fit,
    linkcolor=blue,
    citecolor=blue,
    urlcolor=blue}

\def\be{\begin{equation}}
\def\ee{\end{equation}}
\def\ba{\begin{eqnarray}}
\def\ea{\end{eqnarray}}

\def\lf{\left}
\def\rt{\right}

\begin{document}

\title{Supermassive primordial black holes in multiverse:
for nano-Hertz gravitational wave and high-redshift JWST galaxies}

\author{Hai-Long Huang$^{1}$\footnote{huanghailong18@mails.ucas.ac.cn}}
\author{Yong Cai$^{2}$\footnote{caiyong@zzu.edu.cn}}
\author{Jun-Qian Jiang$^{1}$\footnote{jiangjq2000@gmail.com}}
\author{Jun Zhang$^{3,4}$\footnote{zhangjun@ucas.ac.cn}}
\author{Yun-Song Piao$^{1,3,5,6}$\footnote{yspiao@ucas.ac.cn}}

\affiliation{$^1$ School of Physical Sciences, University of
Chinese Academy of Sciences, Beijing 100049, China}

\affiliation{$^2$ School of Physics and Microelectronics, Zhengzhou University, Zhengzhou, Henan 450001, China}

\affiliation{$^3$ International Centre for Theoretical Physics Asia-Pacific, University of Chinese Academy of Sciences, 100190 Beijing, China}

\affiliation{$^4$ Taiji Laboratory for Gravitational Wave Universe (Beijing/Hangzhou), University of Chinese Academy of Sciences, 100049 Beijing, China}

\affiliation{$^5$ School of Fundamental Physics and Mathematical
    Sciences, Hangzhou Institute for Advanced Study, UCAS, Hangzhou
    310024, China}

\affiliation{$^6$ Institute of Theoretical Physics, Chinese
    Academy of Sciences, P.O. Box 2735, Beijing 100190, China}

\begin{abstract}

Recently, observational hints for supermassive black holes have
been accumulating, which has inspired ones to wonder: Can
primordial black holes (PBHs) be supermassive, in particular with
the mass $M\gtrsim 10^{9}M_\odot$? A supercritical bubble (with an
inflating baby universe inside it) that nucleated during inflation
can develop into a PBH in our observable Universe. Here, we find
that when the inflaton slowly passes by a neighboring vacuum, the
nucleating rate of supercritical bubbles would inevitably attain a
peak, so the mass distribution of multiverse PBHs, and the mass of
peak can be up to $M\gtrsim 10^{11}M_\odot$. Thus our mechanism
naturally provides a primordial origin of supermassive BHs.

\end{abstract}

\maketitle

\section{Introduction}

In past years, the cosmological implications of PBHs
\cite{Zeldovich,Hawking:1971ei,Carr:1974nx}, which might be
responsible for dark matter and LIGO-Virgo gravitational wave (GW)
events \cite{Bird:2016dcv,Clesse:2016vqa,Sasaki:2016jop}, have
been intensively studied, e.g., in
Refs.~\cite{Sasaki:2018dmp,Carr:2020gox,Carr:2023tpt}. However, it
has still been interesting to ask: Can PBHs be supermassive? In
particular can the mass of PBHs reach $M\gtrsim
10^{9}M_\odot$~\cite{Carr:2020erq}?

The origin of supermassive PBHs might have to be related with a
period of inflation, since only inflation can stretch sub-horizon
inhomogeneities to the scale that supermassive PBHs need. It has
been widely thought that massive PBHs can be sourced by very large
inflationary perturbations
\cite{Carr:1993aq,Ivanov:1994pa,Garcia-Bellido:1996mdl,Kawasaki:1997ju,Yokoyama:1998pt},
$\delta\rho/\rho\gtrsim 0.1$. However, current CMB spectral
distortion observations have ruled out such a significant
enhancement of the amplitudes of primordial perturbations (if the
perturbations are Gaussian)\footnote{It seems that one needs to
consider a scenario with highly non-Gaussian primordial
perturbations to create supermassive PBHs,
e.g.\cite{Nakama:2016kfq,Hasegawa:2017jtk,Kawasaki:2019iis,Kitajima:2020kig,Atal:2020yic},
however, see also \cite{Shinohara:2021psq,Shinohara:2023wjd}.} on
$k\lesssim 10^4$Mpc$^{-1}$ scale, and hence PBHs with the mass $M
> 10^4M_\odot$
\cite{Nakama:2017xvq}, see also
Refs.~\cite{DeLuca:2021hcf,DeLuca:2022bjs,Franciolini:2023pbf}.

It has also been well-known that a supercritical bubble (with an
inflating baby universe inside it) that nucleated during inflation
can develop into a PBH in our observable Universe
\cite{Garriga:2015fdk}. In the corresponding multiverse
scenario\footnote{This multiverse PBH scenario is a reminiscent of
the well-known eternally inflating multiverse
\cite{Vilenkin:1983xq,Linde:1986fd}.}, after inflation ended, the
supercritical bubble will connect to our Universe through a
wormhole, and eventually we would see a PBH after the wormhole
pinched off, see also
Refs.~\cite{Deng:2016vzb,Deng:2017uwc,Deng:2020mds,Wang:2018cum,He:2023yvl}
for further investigations. However, the mass distribution of such
multiverse PBHs is $\propto {1\over M^{1/2}}$
\cite{Garriga:2015fdk,Deng:2017uwc}, and thus is negligible at
supermassive band $M\gtrsim 10^{9}M_\odot$, see also
Ref.~\cite{Kusenko:2020pcg}.

Recently, observational hints for supermassive black holes have
been accumulating. The evidences for a nano-Hertz stochastic GW
background have been found with PTA
\cite{NANOGrav:2023gor,Xu:2023wog,Reardon:2023gzh,Antoniadis:2023ott},
which might be interpreted with a population of $M\gtrsim
10^9M_\odot$ supermassive BH binaries
\cite{NANOGrav:2023hfp}\footnote{An alternative might be
inflationary primordial GW
\cite{NANOGrav:2023hvm,Vagnozzi:2023lwo}.}, see also
\cite{Ellis:2023dgf}. Supermassive black holes with $M\sim
10^8-10^{10}M_\odot$ are believed to sit at the center of galaxies
observed at redshifts $z\gtrsim 6$, which has still been a
challenge to the standard astrophysical accretion
models~\cite{Volonteri:2010wz,Volonteri:2021sfo}. The observations
with JWST have also discovered lots of early supermassive galaxies
($M\gtrsim 10^{10}M_\odot$) at higher redshift $z\sim 10$, which
is discordant with the $\Lambda$CDM model
\cite{Boylan-Kolchin:2022kae}, but might be explained with
$\Lambda$CDM+supermassive black holes
\cite{Liu:2022bvr,Hutsi:2022fzw}. Currently, it has been a
consensus that supermassive black holes can be either sourced by
seed-like PBHs as massive as $\sim
10^3M_\odot$~\cite{Duechting:2004dk,Serpico:2020ehh}, which but
must acquire sufficient accretion, or supermassive PBHs by birth,
which might be most natural.

However, it is still unclear how supermassive PBHs can form. In
this work, we present such a mechanism. It is found that in a
slow-roll inflation model with multiple neighboring metastable
vacua, the mass distribution of multiverse PBHs formed by
supercritical bubbles that nucleated during inflation would
inevitably have a multi-peaks spectrum, and the peak mass can
reach $M \gtrsim 10^{11}M_\odot$. Thus, our multiverse PBHs can
naturally serve as supermassive BHs needed to explain nano-Hertz
GW and supermassive JWST galaxies.

\section{Our multiverse PBH model}

\begin{figure}[t]
{\includegraphics[width=7cm]{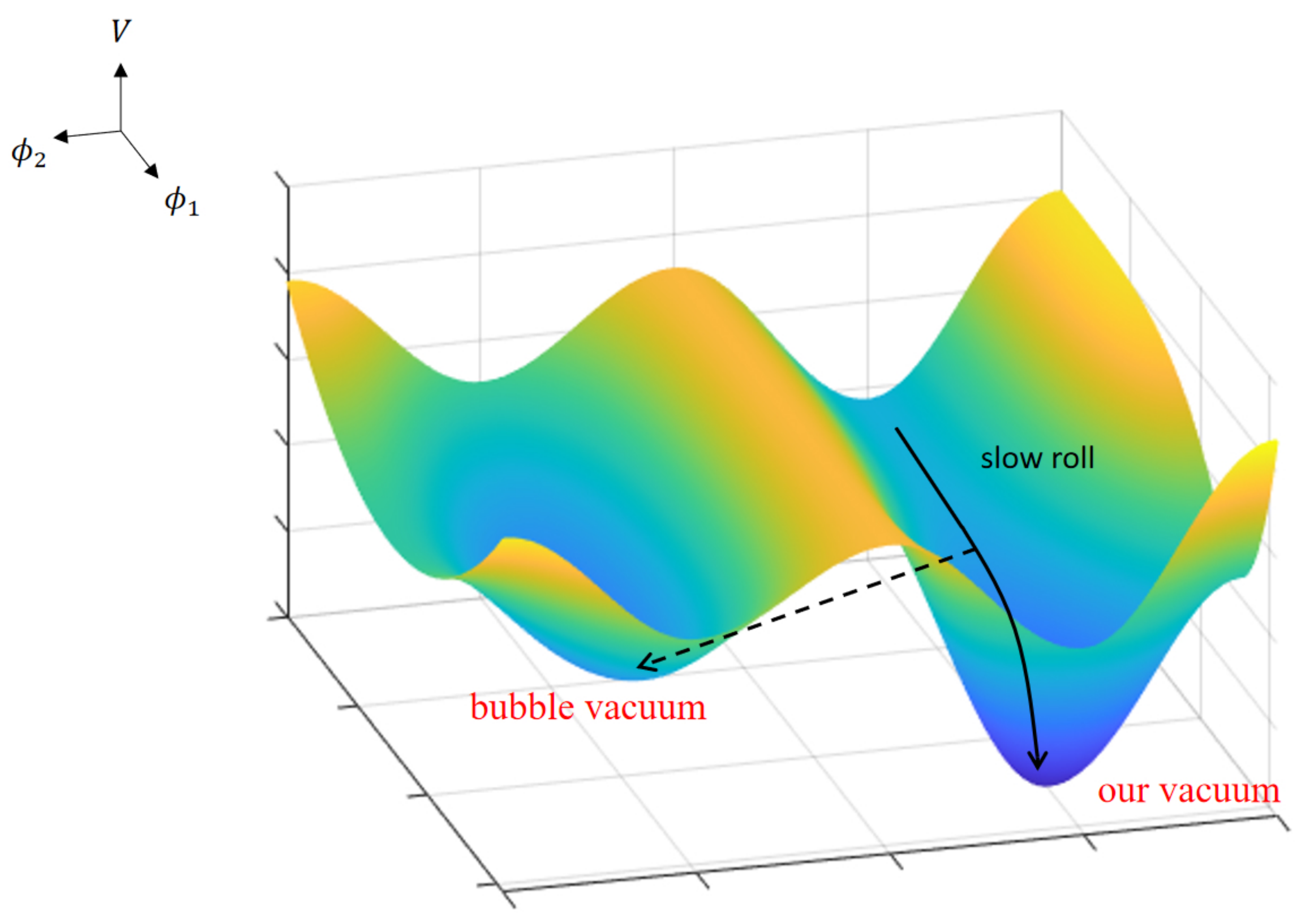}}
    \quad
{\includegraphics[width=7cm]{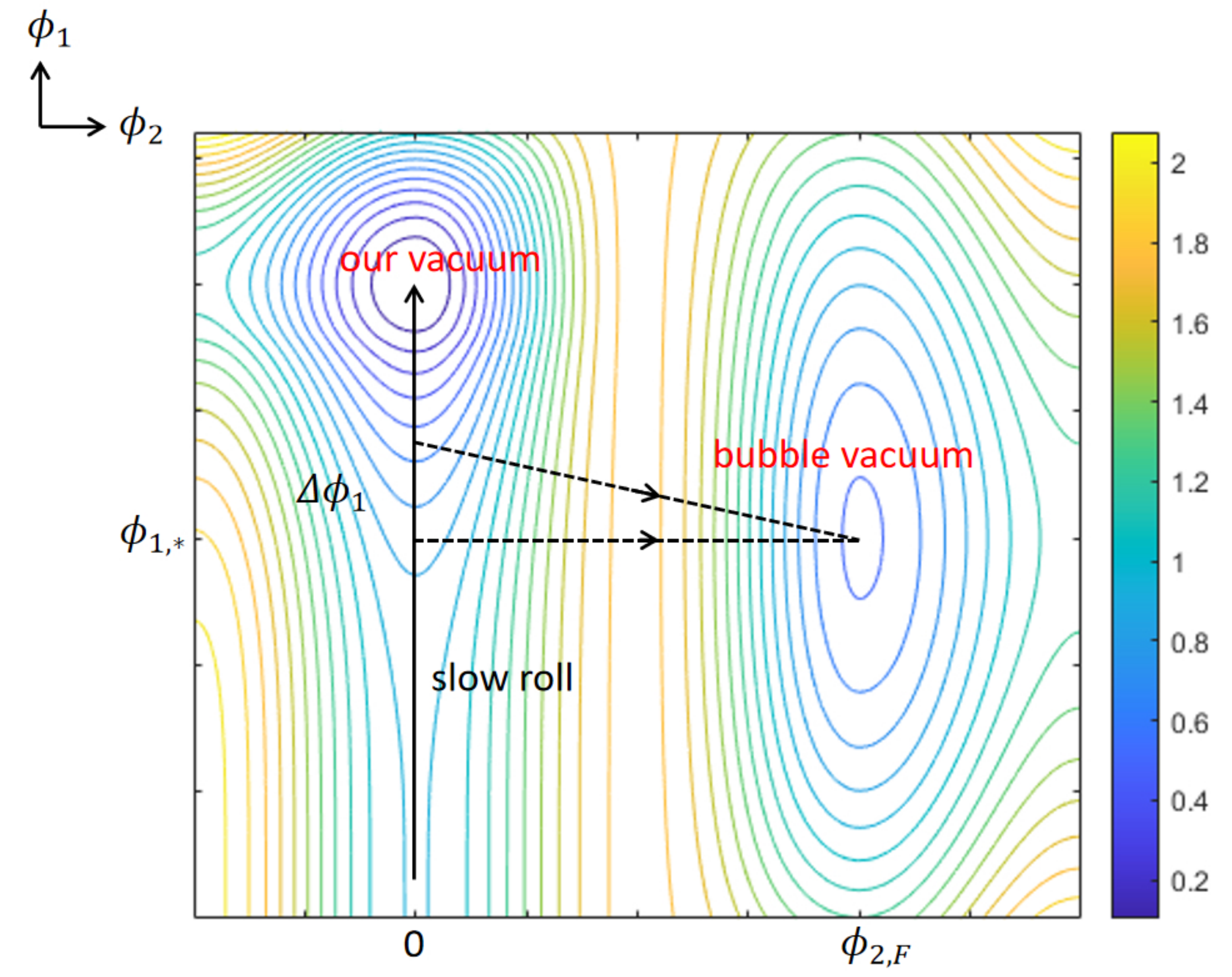}}
    \caption{\textbf{A 2D potential and its contour diagram}. Initially, the
inflaton $\phi_1$ rolls along its potential at $\phi_2=0$, and the
nucleating rate of vacuum bubbles with a neighboring vacuum at
$\phi_2=\phi_{2,F}$ and $\phi_1=\phi_{1,*}$ is highly suppressed.
And only when $\phi_1\simeq \phi_{1,*}$, the bubbles will just
nucleate with largest rate. }
    \label{fig:twofieldpotential}
\end{figure}

In our phenomenological model, see Fig.\ref{fig:twofieldpotential}
for $V(\phi_1,\phi_2)$, the inflaton $\phi_1$ slowly rolls along
its potential $V_{inf}(\phi_1)$ at which $\phi_2=0$, and a
neighboring vacuum with $V_b<V_{inf}$ is at $\phi_2=\phi_{2,F}$
and $\phi_1=\phi_{1,*}$. It is expected that only when $\phi_1$
rolls to the vicinity of $\phi_{1,*}$ the nucleating rate of the
bubble with $V=V_b$ is maximized.

In the thin-wall regime, the bubble nucleating rate per Hubble
spacetime volume $1/H_i^4$ in the inflating background (the Hubble
parameter is $H^2_i={8\pi V_{inf}(\phi_1)\over 3M_P^2}$) is
\cite{Garriga:2015fdk} \be \lambda\sim e^{-B}, \quad \text{with} \quad
B={2\pi^2 \sigma\over H^3_i},\label{B}\ee where \be
\sigma=\int_{path}
\sqrt{2\left(V(\phi_1,\phi_2)-V_b\right)}\lf(\sum_{i=1,2}d\phi_i^2\rt)^{1/2}\label{sigma}\ee
is the wall tension with the ``path" representing the
``least-$\sigma$" path~\cite{Ahlqvist:2010ki}\footnote{The
nucleating rate of bubbles in multiple-field space has been also
investigated in
e.g., Refs.~\cite{Kusenko:1995jv,Konstandin:2006nd,Masoumi:2016wot,Espinosa:2018szu}.}.
Here, we will not calculate $\sigma$ exactly, but consider such an
approximation (which helps to highlight the essential how
$\lambda$ is affected by the roll of inflaton), i.e., when inflaton
passed through $\phi_{1,*}$, we have $d\phi_1\simeq
\Delta\phi_1=\phi_1-\phi_{1,*}$ and $d\phi_2\simeq\phi_{2,F}$,
thus
\be \sigma \thickapprox \lf(1+{(\Delta\phi_1)^2\over
\phi_{2,F}^2}\rt)^{1/2}\int_0^{\phi_{2,F}}
\sqrt{2\left(V(\phi_{1,*},\phi_2)-V_b\right)}d\phi_2.
\label{eq:sigma} \ee According to (\ref{B}), we have \be
B\thickapprox B_*\lf(1+{(\Delta\phi_1)^2\over
\phi_{2,F}^2}\rt)^{1/2},\quad \text{with} \quad B_*={{2\pi^2 \over H_i^3}
\int_0^{\phi_{2,F}}
\sqrt{2\left(V(\phi_{1,*},\phi_2)-V_b\right)}d\phi_2},
\label{BM}\ee where the effect of the rolling velocity of inflaton
on $\lambda$ has been imprinted in $\Delta\phi_1=\int {{\dot
\phi}_1\over H}d{\cal N}$, with ${\cal N}=\int Hdt=\int
{Hd\phi\over {\dot\phi}}$ being the e-folding number before inflation
ended.


\begin{figure}[t]
{\includegraphics[width=4cm]{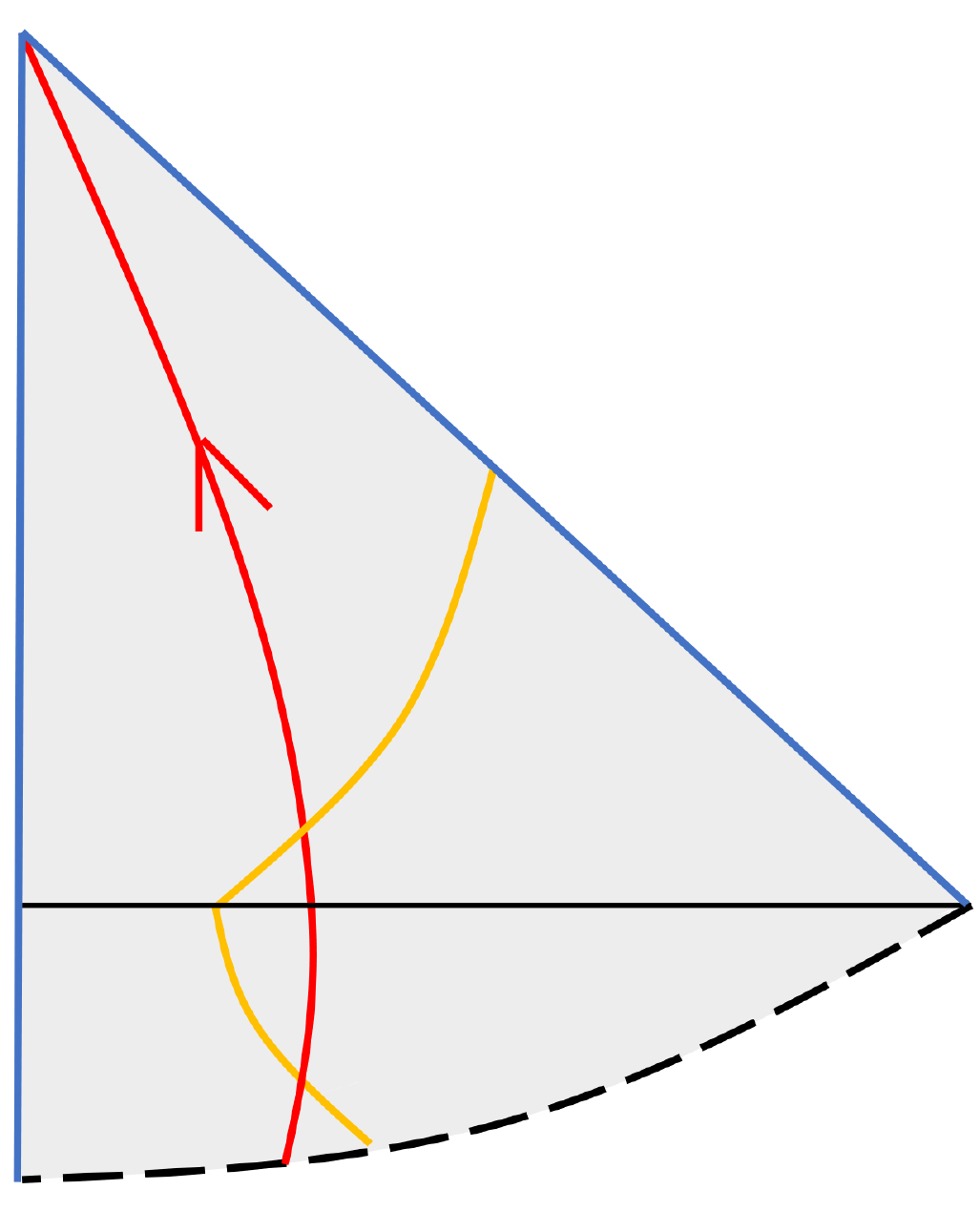}}
    \qquad \qquad
{\includegraphics[width=8cm]{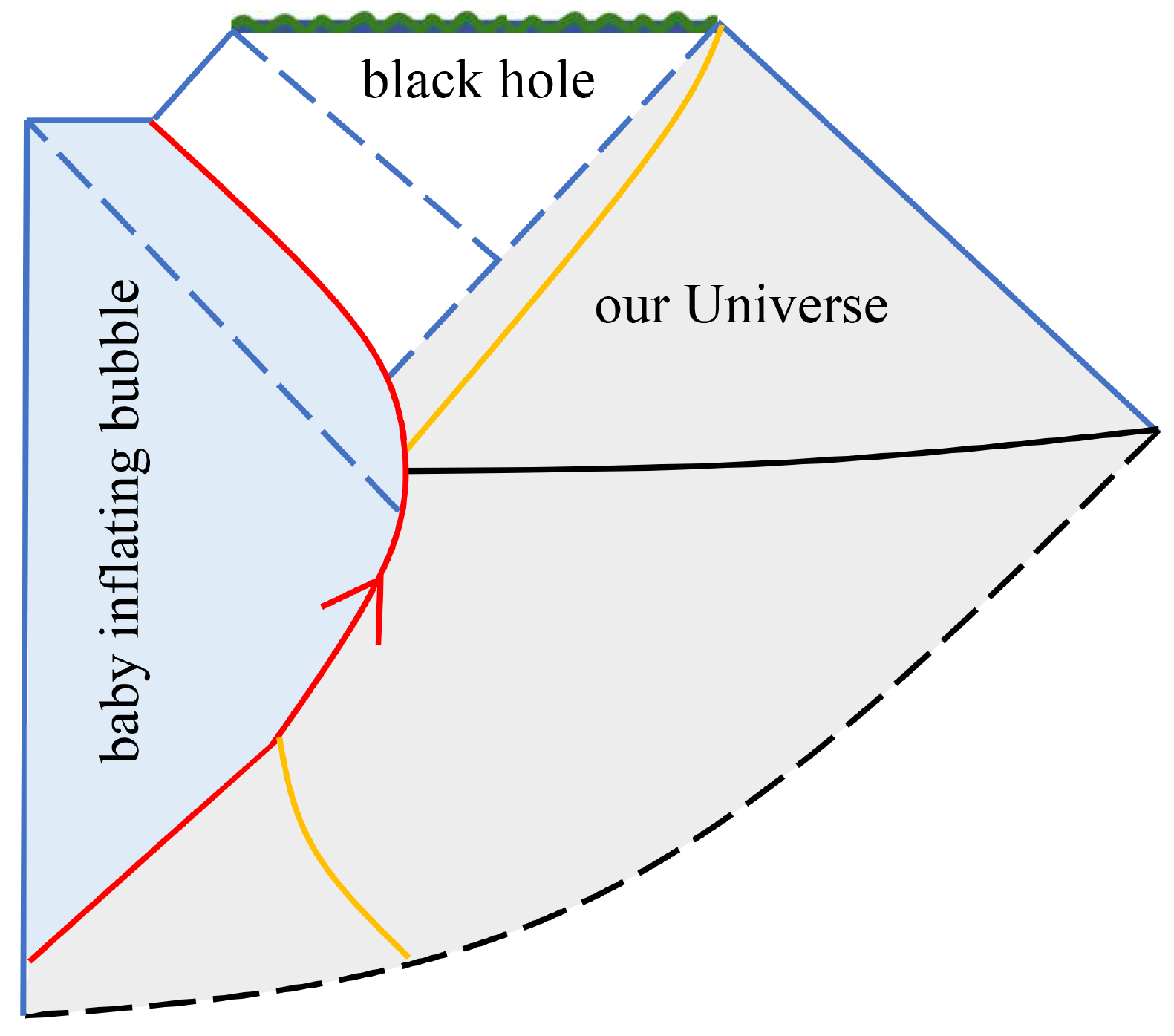}}
\caption{\textbf{Right
larger panel: Penrose diagram of inflation followed by a radiation
era with multiverse PBH.} The red curve is the comoving bubble
wall, the yellow curves are the comoving Hubble horizon, and the
black dashed and solid curves represent when the inflation started
and ended, respectively. The bubble that nucleated during
slow-roll inflation will evolve to be supercritical, i.e.
$r\gtrsim 1/H_b$, its interior contains a baby inflating universe
(blue region). The bubble entered into the horizon of our
observable Universe at $t=t_H$, after which it will be hidden
behind the horizon of a PBH. \textbf{Left panel (A contrast):
Penrose diagram of inflation followed by a radiation era.} The red
curve is the comoving primordial perturbation. }
    \label{fig:Penrose}
\end{figure}

The bubble after its nucleating will rapidly expand. The inflation
ended at $t=t_i$, at which time the number density of bubbles is \be
dn(t_i)=\lambda {dr_{i}\over \lf(r_i+H^{-1}_i\rt)^4},
\label{nti}\ee ($r_i={e^{\cal N}\over H_i}\gg 1/H_i$ is the radius
of bubble at $t_i$), and the energy of inflaton is rapidly
converted to that of radiation, so $V\sim 0$ outside the bubble is
completely negligible.

In Fig.\ref{fig:Penrose}, we present the full Penrose diagram for
the supercritical bubble \footnote{The supercritical bubble refers
to the bubble with its radius larger than the Hubble length of
spacetime inside bubble, i.e. $r\gtrsim {1\over H_b}$, where
$H^2_b={8\pi\over 3M_P^2}V_b$.} evolving into a PBH. In
corresponding spacetime, the interior of supercritical bubble
contains a baby inflating universe, which is connected to the
exterior through a wormhole being closed. And eventually a PBH
will come into being in our observable Universe.\footnote{In
earlier Refs.~\cite{Maeda:1981gw,Kodama:1981gu,
Farhi:1986ty,Blau:1986cw}, the scenario with baby inflating
universes inside black holes in de Sitter or asymptotically flat
spacetime have been explored.}

In light of the causality, the region affected by the
Schwarzschild radius of the PBH cannot exceed the Hubble radius of
the parent Universe $1/H\simeq t_{H}$. During the radiation era,
the bubble will expand comovingly with $a(t)=({t\over t_i})^{1/2}$,
until its physical radius $r=a(t_H)r_i\sim t_{H}$ (so
$t_H=H_ir^2_i$), after which the bubble will be hidden behind the
horizon of a PBH with mass $M\sim M_p^2t_{H}$, so \be M\sim
H_ir_i^2M_P^2={M_P^2\over H_i}e^{2{\cal N}}.\ee The dark matter
density is $\rho_{DM}(t)\sim {M_P^3\over t^{3/2}{\cal
M}_{eq}^{1/2}}$, where ${\cal M}_{eq}\sim 10^{17}M_{\odot}$.
Thus the mass distribution of such PBHs is~\cite{Garriga:2015fdk,Deng:2017uwc} \be f(M)={1\over
\rho_{DM}(t)}\lf(M^2{dn\over dM}\rt)\sim \lambda\lf({{\cal
M}_{eq}\over M}\rt)^{1/2}. \label{M1}\ee According to (\ref{BM}),
$f(M)\propto {e^{-B_*}\over M_*^{1/2}}$ is maximized\footnote{In
Ref.\cite{Garriga:2015fdk}, $\lambda\sim const.$, so $f(M)\propto
{1\over M^{1/2}}$.} at $\phi_{1}=\phi_{1,*}$ (equivalently
$M_*={M_P^2\over H_i}e^{2{\cal N}_*}$).



\begin{figure}[tb]
    {\includegraphics[width=7cm]{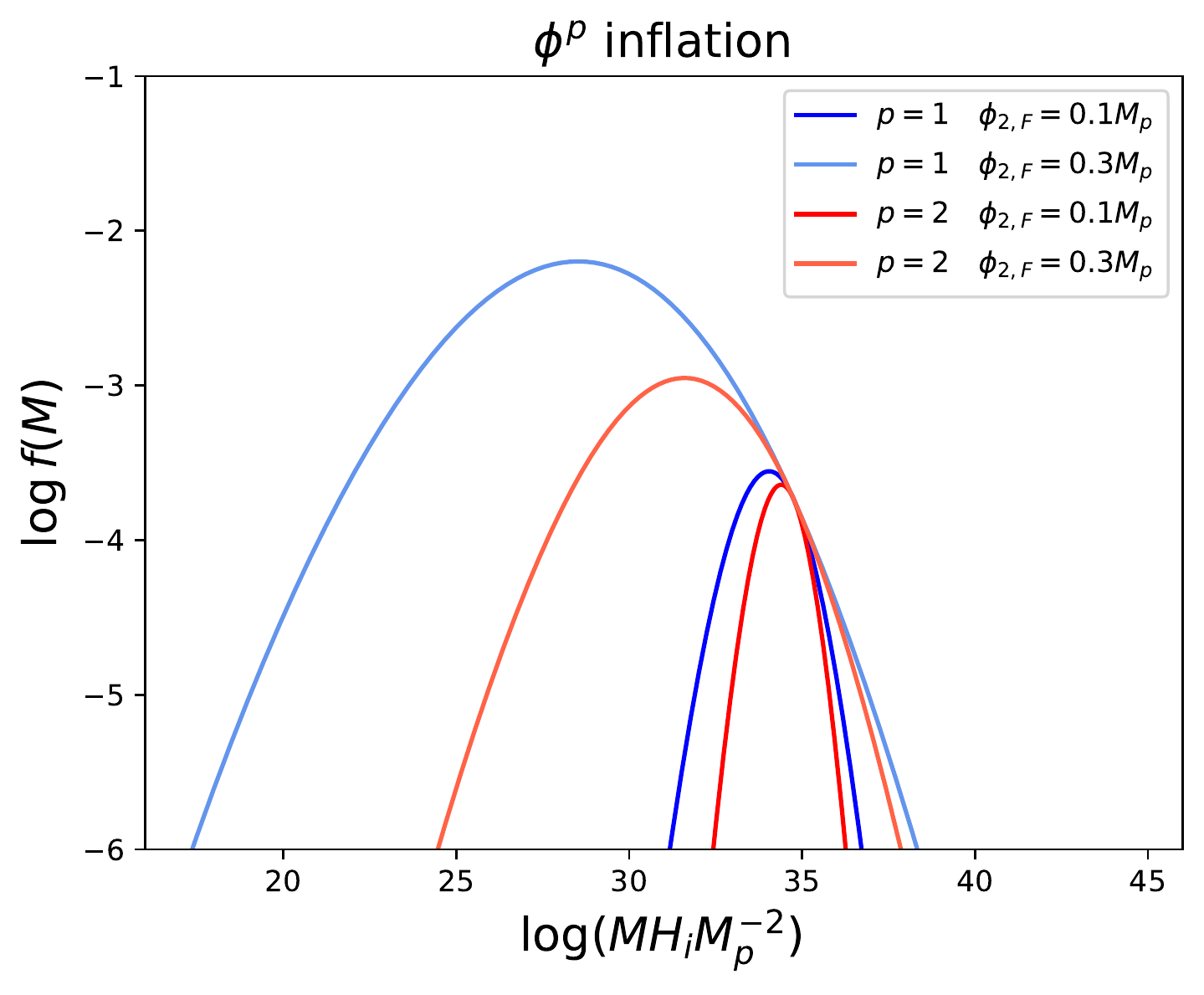}}
    \quad
    {\includegraphics[width=7cm]{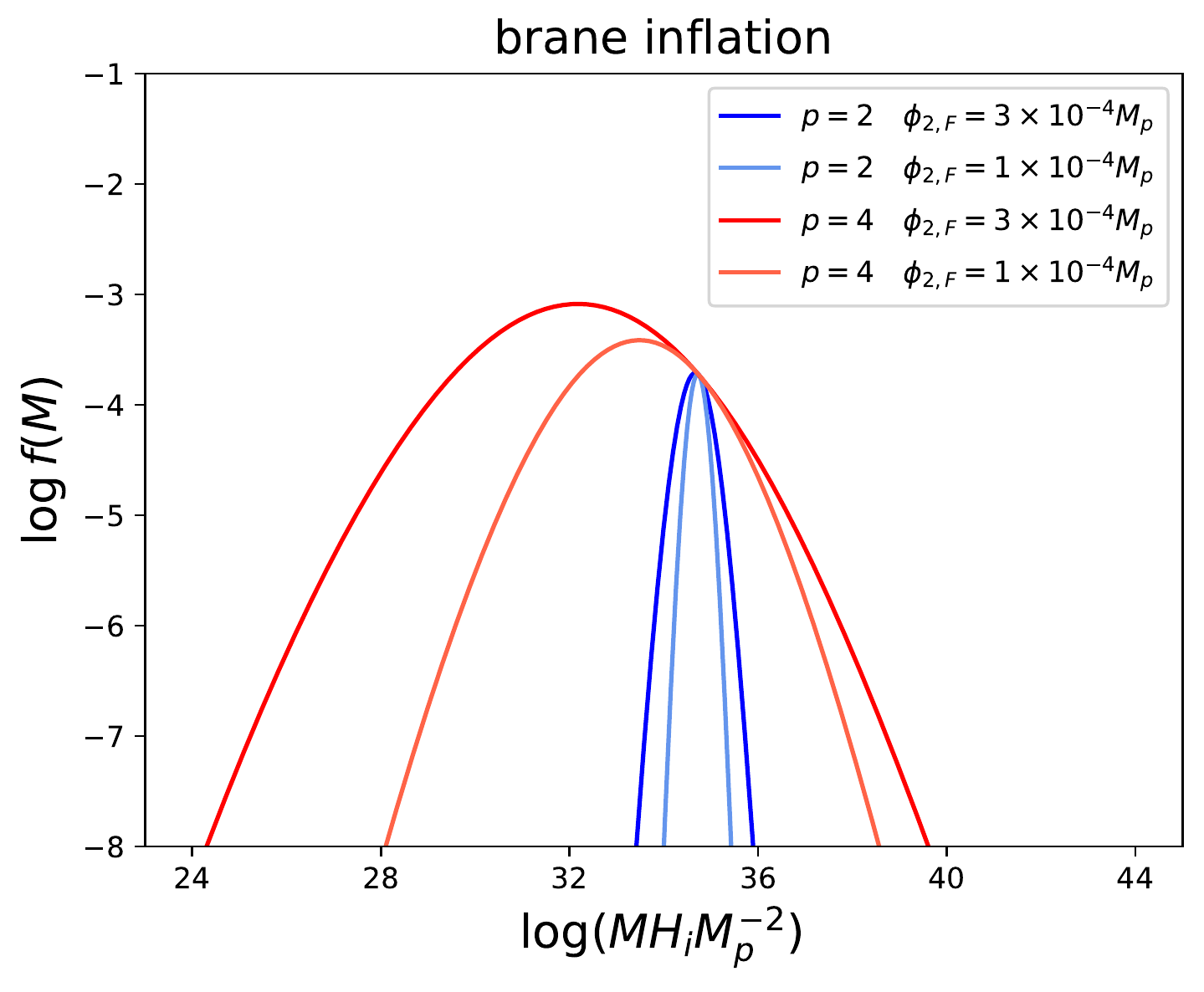}}
\caption{\textbf{The mass spectra $f(M)$ for different models of
slow-roll inflation.} We set $\mathcal{N}_*=40$, and specially for
KKLT brane inflation, $\mu=\phi_{2,F}\ll M_P$. }
    \label{fig:phininflaion}
\end{figure}

It is significant to see the resulting $f(M)$ in different
slow-roll inflation models. As a simple example, we consider
$V_{inf}(\phi_1)\sim \phi_1^p$. In large-${\cal N}$ limit, \be
{\cal N}\simeq {4\pi\over pM_P^2} \phi_1^2.\ee Here, $p=2$
corresponds to the chaotic inflation
\cite{Linde:1983gd,Kaloper:2008fb}\footnote{Recently, based on
$\Lambda$CDM model the Planck collaboration has obtained
$n_s\approx 0.965$ \cite{Planck:2018jri}. However, $n_s=1$
($n_s-1\sim -{\cal O} (0.001)$) is also observationally allowed
\cite{Ye:2020btb,Ye:2021nej,Jiang:2022uyg,Jiang:2022qlj} in light
of resolution of recent Hubble tension. Thus chaotic inflation
might be still consistent in its hybrid extension
\cite{Kallosh:2022ggf,Ye:2022efx}}, $p=2/3,1,4/3$ correspond to
the monodromy inflation
\cite{Silverstein:2008sg,McAllister:2008hb}. Thus with \eqref{BM},
we have \be B\thickapprox B_*\sqrt{1+{pM_P^2\over
4\pi\phi_{2,F}^2}\lf(\mathcal{N}^{1/2}-\mathcal{N}_*^{1/2}\rt)^2},\label{phip}\ee
which suggests \ba f(M)&\sim & e^{-B({\cal N})}\lf({{\cal
M}_{eq}\over M}\rt)^{1/2}\nonumber\\ &=&\lf({{\cal M}_{eq}H_i\over
M_P^2}\rt)^{1/2}\exp\lf({-B_*\sqrt{1+{pM_P^2\over
4\pi\phi_{2,F}^2}\lf(\mathcal{N}^{1/2}-\mathcal{N}_*^{1/2}\rt)^2}-{\cal
N}}\rt), \ea where $\mathcal{N}=\ln\sqrt{MH_i\over M_P^2}$. The
results with ${\cal N}_*=40$ are shown in
Fig.~\ref{fig:phininflaion}. Different $p$ and $\phi_{2,F}$
(``shortest paths" to a neighboring vacuum) result in different
profiles of $f(M)$.


In our multiverse PBH model, we can have $\epsilon=-{\dot
H}/H^2\sim 0.01$ on all scales (outside the bubbles) for $\phi^p$
inflation. However, it must be mentioned that in
primordial-perturbation-sourced PBH models, a large amplitude
$\delta\rho/\rho\gtrsim 0.1$ of primordial scalar perturbations
requires that the standard slow-roll evolution, $\epsilon\sim
0.01$, of inflaton must be broke on corresponding PBH scales,
e.g., so-called ultra-slow roll\footnote{In such single-field
inflation model, a recent dispute is whether the enhanced
small-scale perturbations might lead to large one-loop correction
for perturbations on CMB scales or not,
e.g.\cite{Kristiano:2022maq,Riotto:2023hoz,Kristiano:2023scm,Fumagalli:2023hpa}.}.


It is also interesting to consider KKLT brane inflation,
$V_{inf}(\phi_1)\sim 1-({\mu\over \phi_1})^p$
\cite{Kachru:2003sx,Kallosh:2018zsi}, where $p=1,2,3,4$. In such
models, for $\mu\ll M_p$, we have \be {\cal N}= {8\pi\mu^2\over
p(p+2)M_P^2}\lf({\phi\over \mu}\rt)^{p+2}.\ee Thus with
(\ref{BM}), we have \be B\thickapprox B_*\sqrt{1+{\cal
A}{\mu^{2p/(p+2)}M_P^{4/(p+2)}\over
\phi_{2,F}^2}\lf(\mathcal{N}^{1\over p+2}-\mathcal{N}_*^{1\over
p+2}\rt)^2},\label{Bbrane}\ee where ${\cal A}=\lf({p(p+2)\over
8\pi}\rt)^{2/(p+2)}$. The corresponding results are also shown in
Fig.\ref{fig:phininflaion}.

\section{Discussion}

In this work, we found that if the inflaton passed by a
neighboring vacuum, the mass spectrum of multiverse PBHs formed by
the supercritical bubbles that nucleated during inflation would
attain a peak. It is usually expected that inflation happened at
$V_{inf}\sim (10^{15}$Gev$)^4$ and lasts for ${\cal N}\thickapprox
60$ efolds, thus if the bubbles nucleated at $10 \lesssim {\cal
N}_*\lesssim 50$, the mass peak will be at \be
10^{-22}M_\odot\lesssim M_*={M_P^2\over H_i}e^{2{\cal
N}_*}\lesssim 10^{11}M_\odot. \label{MBH}\ee In
Ref.~\cite{Liu:2022bvr}, it has been shown that early supermassive
galaxies observed by JWST can be explained with supermassive ($M
\gtrsim 10^9M_\odot$) BHs that make up a small fraction ($\sim
10^{-6}-10^{-3}$) of dark matter. According to (\ref{MBH}), our
multiverse PBHs can be supermassive up to $10^{11}M_\odot$, which
thus can naturally serve as such supermassive BHs.

The nano-Hertz stochastic GW background recently detected might be
interpreted with a population of $M\gtrsim 10^9M_\odot$
supermassive BH binaries, but such a explain seems to have a mild
tension with NANOGrav 15-Year data
\cite{NANOGrav:2023hfp,Ellis:2023dgf}. However, the result is
actually dependent on the origin and evolution mechanism of
supermassive BHs. In this sense, current tension might be just a
reflection of the primordial origin of supermassive BHs.

\begin{figure}[tb]
{\includegraphics[width=12cm]{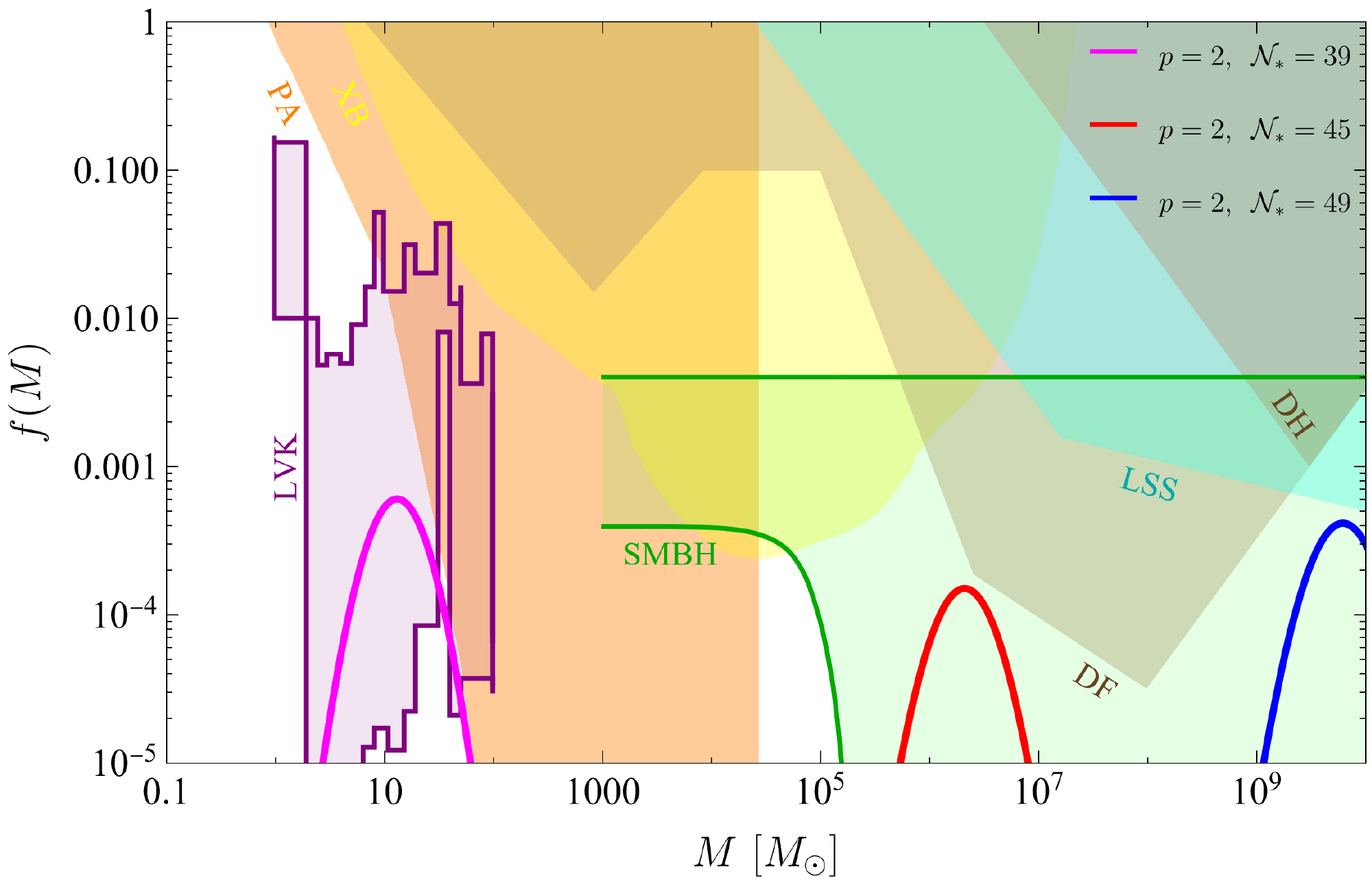}}
\caption{\textbf{The
multi-peaks mass spectrum of our multiverse PBHs.} In our model,
the inflaton slowly rolled through near different vacua when about
${\cal N_*}=39, 45, 49$. As a result, the mass distributions of
multiverse PBHs sourced by the supercritical bubbles will peak at
different mass $M\sim 10$ (purple curve), $10^6$ (red curve),
$10^{10}M_\odot$ (blue curve), respectively. According to Fig.38
in Ref.\cite{Carr:2023tpt}, we also show current constraints on
massive PBHs.} \label{Fig.SMBH}
\end{figure}

Though we have presented a mechanism for the origin of
supermassive PBHs, it can be also expected that in a string theory
landscape, the slow-roll path of inflaton would be accompanied
with more than one neighboring vacua, so that our multivers PBHs
would have a multi-peaks mass spectrum, which might simultaneously
account for current LIGO-Virgo GW events, e.g., see
Ref.~\cite{He:2023yvl}, supermassive BHs ($M\sim 10^{9}M_\odot$),
see Fig.\ref{Fig.SMBH}, and other observations hinting PBHs
\cite{Carr:2023tpt}. Thus it is interesting to further study the
implications of recent observations at corresponding mass band,
such as LIGO-Virgo-KAGRA GWTC events and early supermassive JWST
galaxies, on our multivers PBHs.

Here, it is significantly noted that the multi-peaks mass
spectrum of our multiverse PBHs encoded the information of not
only slow-roll inflation but also the string vacua
\cite{Susskind:2003kw,Kachru:2003aw}, which might be a potential
probe to relevant issues and worthy of explorations. It is also
interesting to ask if such multiverse PBHs help to resolve the
information paradox \cite{Arkani-Hamed:2007ryv} of eternally
inflating spacetime, e.g., see Ref.~\cite{Piao:2023vgm}. Though our
simplified calculation might capture the essential how the
nucleating rate is affected by the roll of inflaton, it is
necessary to calculate it in a well-motivated string theory
landscape, which needs to be further investigated to better
understand the origin of supermassive PBHs.

\section*{Acknowledgments}
YSP is supported by NSFC, No.12075246 and the Fundamental Research
Funds for the Central Universities.
Y. C. is supported in part by the National Natural Science Foundation
of China (Grant No. 11905224), the China Postdoctoral Science
Foundation (Grant No. 2021M692942) and Zhengzhou University (Grant No. 32340282).
J. Z. is supported by the scientific research starting grants from University of Chinese Academy of Sciences (Grant No.~118900M061), the Fundamental Research Funds for the Central Universities (Grant No.~E2EG6602X2 and Grant No.~E2ET0209X2) and the National Natural Science Foundation of China (NSFC) under Grant No.~12147103.


 \end{document}